\documentclass[%
reprint, superscriptaddress,
showpacs,preprintnumbers,
dvipsnames,svgnames,table,
 amsmath,amssymb,
 aps]{revtex4-1}
\usepackage{amsmath}
\usepackage{amssymb}
\usepackage{graphicx}
\usepackage[caption=false]{subfig}
\usepackage{float}
\usepackage[left=25mm,right=25mm,top=25mm,bottom=25mm]{geometry}
\usepackage[usenames,dvipsnames]{xcolor}
\usepackage{enumitem}
\usepackage{epstopdf}
\usepackage{hyperref}
\hypersetup{
    colorlinks=true,
    linkcolor=blue,
    }
\usepackage[authormarkup=none]{changes}
\definechangesauthor[name={Stefan}, color=purple]{n}
\definechangesauthor[name={Stella}, color=blue]{s}
\definechangesauthor[name={Valerio}, color=red]{v}
\definechangesauthor[name={Angeline}, color=orange]{a}
\definechangesauthor[name={Caiyu}, color=green]{c}

\newcommand{\ket}[1]{\left|#1\right\rangle}
\newcommand{\bra}[1]{\left\langle#1\right|}

\newcommand{\ba}{\begin{eqnarray}}\newcommand{\ea}{\end{eqnarray}}
\newcommand{\ban}{\begin{eqnarray*}}\newcommand{\ean}{\end{eqnarray*}}

\begin{document}
%\title{Stability of the second laws of thermal operations under relaxations of the framework}

\title{Almost thermal operations: inhomogeneous reservoirs}

\author{Angeline Shu}
 \affiliation{Department of Physics, National University of Singapore, 2 Science Drive 3, Singapore 117542.}
 %\affiliation{Centre for Quantum Technologies, National University of Singapore, 3 Science Drive 2, Singapore 117543.}
\author{Yu Cai}
\affiliation{D\'epartement de Physique Appliqu\'ee, Universit\'e de Gen\'eve, 1221 Gen\'eve, Switzerland.}
\author{Stella Seah}
 \affiliation{Department of Physics, National University of Singapore, 2 Science Drive 3, Singapore 117542.}
\author{Stefan Nimmrichter}
 %\affiliation{Centre for Quantum Technologies, National University of Singapore, 3 Science Drive 2, Singapore 117543.}
 \affiliation{Max Planck Institute for the Science of Light, Staudtstra{\ss}e 2, 91058 Erlangen.}
\author{Valerio Scarani}
 \affiliation{Department of Physics, National University of Singapore, 2 Science Drive 3, Singapore 117542.}
 \affiliation{Centre for Quantum Technologies, National University of Singapore, 3 Science Drive 2, Singapore 117543.}
\begin{abstract}

The resource theory of thermal operations explains the state transformations that are possible in a very specific thermodynamic setting: there is only one thermal bath, auxiliary systems can only be in the corresponding thermal state (free states), and the interaction must commute with the free Hamiltonian (free operation). In this paper we study the mildest deviation: the reservoir particles are subject to inhomogeneities, either in the local temperature (introducing resource states) or in the local Hamiltonian (generating a resource operation). For small inhomogeneities, the two models generate the same channel and thus the same state transformations. However, their thermodynamics is significantly different when it comes to work generation or to the interpretation of the ``second laws of thermal operations''.
 
\end{abstract}

\maketitle

\section{Introduction}

Foundationally, thermodynamics is a theory of states and their transformations. In quantum information science, the same can be said for entanglement theory. This analogy was discussed very early \cite{Opp2002,Horo2002}, and has later resulted in the development of the broad framework of resource theories. Among those, the \textit{resource theory of thermal operations} is a formalisation of the thermodynamics of systems in contact with thermal baths \cite{Brandao_2013,Gour_2015,Goold_2016}. The lack of resources is described by what can be achieved with a single thermal bath at temperature $T$ (because with two different temperatures one can run an engine). Specifically, the \textit{free states} are the thermal states $\tau$ at temperature $T$, and the \textit{free operations} $U$ are those that conserve the total energy. Both notions are defined with respect to a reference Hamiltonian, usually taken as $H=H_S+H_R$ where $S$ indicate the system and $R$ a reservoir of auxiliary systems. Then, thermal states read $\tau=\tau_S\otimes\tau_R$ where $\tau_X=e^{-\beta H_X}/Z_X$, $Z_X=\textrm{Tr}(e^{-\beta H_X})$ and $\beta=1/k_{\rm B}T$. An operation represented by the unitary $U$ is a free operation if \ba\label{conserve}
[H,U]=0\,.
\ea
If the system is prepared in the state $\rho$, a \textit{free evolution} (i.e. one that can be achieved without resources) is then of the form
\ba\label{dynamics}
\mathcal{E}[\rho]&=& \textrm{Tr}_{R}\bigl[U (\rho\otimes\tau_{R})U^{\dagger}\bigr]\,.
\ea
Recent studies have addressed the robustness of the framework under modifications of the states \cite{Meer_2017,Mueller_2017,Baumer2019} or of the modelling of the thermal bath \cite{Sparaciari_2017,Scharlau2018,Richens2018}. In this paper, we look at what is arguably the mildest form of deviation: an \textit{inhomogeneous reservoir}. This is a reservoir made of a large number $N$ of systems, whose local parameters deviate randomly from those that would define an exact thermal operation. For this first study, we shall focus on inhomogeneities either in local temperature or in the local Hamiltonian.
%The plan of the paper is as follows. In section \ref{sec:ToyMod}, we introduce the toy model and derive analytical results for the average over several realisations. Section \ref{sec:results} presents and discusses the results obtained for qubit (\ref{sec:qubit}) and qutrit systems (\ref{sec:qutrit}). The physical relevance of the model is discussed in Section \ref{sec:PhyRel} before concluding in Section \ref{sec:conclusion}.

%***************************************************************************************************************************************
%*************************** Section I ****************************
%***************************************************************************************************************************************

\section{The model}
\label{sec:ToyMod}

\subsection{Introducing inhomogeneities}

The system is a qudit, and the reservoir consists of $N$ qudits labelled by $r\in\{1,2,...,N\}$. We work with a Hamiltonian of non-interacting systems
\ba\label{Ham}
H=H_S+H_R&=&g_0 s_z^{(S)}\,+\,\sum_{r=1}^N g_r s_z^{(r)}
\ea where $g_0>0$, the $g_r$ will be discussed later, and $s_z$ is the operator representing the spin in the direction $z$. For every qudit, the eigenstates of $s_z$ for the eigenvalue $ \big(j-\frac{d-1}{2}\big)\hbar$ is denoted by $\ket{j}$ with $j\in\{0,1,...,d-1\}$ --- in particular, $\ket{0}$ is the ground state of $gs_z$ whenever $g>0$.

For simplicity, throughout this work we consider input states of the system $\rho=\sum_jp_j\ket{j}\bra{j}$ that are \textit{diagonal} in the eigenbasis of $H_S$. The qudits of the reservoir are prepared in the thermal state at the local temperature: $\tau_R=\bigotimes_r \tau_r$ with $\tau_r=e^{-\beta_r g_r s_z}/Z_r$.

The inhomogeneous reservoir is described by a configuration $\underline{\delta}_N=(\delta_1,...,\delta_N)$, where $\delta_r$ is the inhomogeneity perceived by the $r$-th qudit of the reservoir. As random variables, we assume that the $\delta_r$ are independent and identically distributed (i.i.d.) with a distribution $G(\delta)$ centered at $\overline{\delta}=0$. We consider two cases: that of \textit{inhomogeneous Hamiltonian} defined by
\ba\label{inH}
g_r=g_0(1+\delta_r)&\textrm{ and }&\,\beta_r=\beta\;\forall r\,;
\ea and that of \textit{inhomogeneous temperature} defined by 
\ba\label{inT}
\beta_r=\beta(1+\delta_r)&\textrm{ and }&g_r=g_0\;\forall r\,.
\ea In the language of resource theories, \eqref{inH} allows for \textit{resource operations} which violate \eqref{conserve}; while \eqref{inT} amounts to considering \textit{resource states} arising from having multiple temperatures. We also note that both inhomogeneities have a clear physical flavor. For instance, if the qudits are magnetic moments, conditions \eqref{inH} may describe the inhomogeneity of the intensity of the local magnetic field, or of the gyromagnetic factor (e.g. through the chemical environment).

Either way, the thermal state $\tau_r$ of each reservoir qudit is
\ba\label{taudelta}
\tau(\delta_r)&=&\frac{e^{-\beta g_0\,(1+\delta_r)s_z}}{\textrm{Tr}(e^{-\beta g_0\,(1+\delta_r)s_z})}\,=\,\sum_jq_{j}(\delta_r)\ket{j}\bra{j}
\ea where $q_{j}(\delta)=\frac{1-a(\delta)}{1-a(\delta)^d}\,a(\delta)^{j}$ with $a(\delta)=e^{-\beta \hbar g_{0}(1+\delta)}$. Clearly $\tau(\delta=0)=\frac{e^{-\beta g_0\,s_z}}{\textrm{Tr}(e^{-\beta g_0\,s_z})}\equiv\tau_S$ the thermal state of the system for $\beta$.

%\footnote{One could also imagine an inhomogeneity in the direction of the local magnetic field: this would be desrcibed by replacing the reservoir terms $g_r s_z^{(r)}$ in \eqref{Ham} with $\vec{g}_r \cdot\vec{s}^{\,(r)}$.} 

\subsection{Interaction: collisional model}

Now we have to discuss the interaction $U$. With the aim of bringing out local inhomogeneities, it is convenient to have the system interact sequentially with each reservoir qudit. In other words, $U=U_{S,N}U_{S,N-1}...U_{S,1}$ is going to be the product of successive two-body interactions, each between the system and one of the reservoir qudits. Such \textit{collisional models} have been used as toy models in several studies of quantum dynamics and thermodynamics, see e.g.~\cite{Valerio_2002,Bruneau2014,Lorenzo_2015,Strasberg_2017,Seah2019,Baumer2019}, although not all thermal operations can be written in this form \cite{Lostaglio2018}. In this paper we assume that all two-body interactions $U_{S,r}$ are given by the partial swap with mixing angle $\theta$:
\ba
U_{S,r}&=&\cos\theta\,\mathbb{I} +i\sin\theta\,\mathcal{S}\label{partialswap}
\ea with $\mathcal{S}$ the swap operator for 2 qudits.
If $g_r=g_0$ for all $r$, then $U$ couples only degenerate eigenstates of $H$ and \eqref{conserve} holds.

\subsection{Dynamics of the system}

In the absence of inhomogeneities ($\tau_r=\tau_S$ for all $r$, that is $\underline{\delta}_N=0$), the dynamics \eqref{dynamics} can be solved analytically for our model. For diagonal input states, the state of the system after interaction with the first $r$ qudits of the reservoir is given by 
\ba
    \rho_{S|r}&=\rho_{S|r-1} \cos^2 \theta + \tau_S \sin^2 \theta \nonumber \\
    &=\tau_S - \left(\tau_S-\rho_{0}\right)\cos^{2r} \theta\;\,. \label{eq:Analytic}
\ea
In particular, the state remains diagonal and converges to the thermal state $\tau_S$ in the limit $N\rightarrow\infty$.

Each configuration $\underline{\delta}_N$ of the inhomogeneities induces a new map on the system. If the inhomogeneities are frozen, the dynamics \eqref{dynamics} defines a contractive map whose fixed point $\rho_{S|\infty}$ is determined by the specific $\underline{\delta}_N$, and there is little more to say. The model is more interesting if $\underline{\delta}_N$ is drawn independently for each use of the channel: then we can study the \textit{ensemble average} over $G(\delta)$. The dynamics commutes with this average: for i.i.d.~inhomogeneities, the reservoir qudits are all prepared in the ensemble-averaged thermal state
\ba\label{ensavg}
\overline{\tau}&=&\int_{-\infty}^{\infty}G(\delta)\tau(\delta)\,\mathrm{d}\delta\,.
\ea Thus the similarly defined ensemble-averaged state of the system at step $r$ is \begin{align}
    \overline{\rho}_{S|r}&= \overline{\tau}- \left(\overline{\tau}-\rho_{0}\right) \cos^{2r}\theta\,.\label{eq:Analytic_Noise}
\end{align} For $d>2$, $\overline{\tau}$ won't be thermal in general. For qubits, $\overline{\tau}$ can be seen as a thermal state for an effective temperature larger than $T$ as a convex combination of density matrices necessarily increases the entropy and hence decreases the purity of the state. Since the occupation of the ground state is smaller in $\overline{\tau}$ than it is in $\tau_S$, not unexpectedly the evolution violates majorisation.

%***************************************************************************************************************************************
%*************************** Section III ****************************
%***************************************************************************************************************************************

\section{Work, heat and first law -- Inhomogeneous hamiltonian}
For each two body interaction between the system and a single reservoir qudit, any change to energy of the combined system and reservoir qudit will be construed as work input due to the presence of interaction, $W=\Delta \mathrm{Tr}[\rho H]$. Here $\rho$ refers to the density matrix of the combined system and reservoir qudit. Since the degrees of freedom of the reservoir are generally inaccessible to us, we can identify any changes of energy of the reservoir alone as heat output, $Q=-\Delta \mathrm{Tr}[\rho_R H_R]$. The net change in the energy of the system alone then obeys the first law by construction, $\Delta U=\Delta\mathrm{Tr}[\rho_S H_S] = Q+W$.

In the following two sections, we look at the statistics of work and heat for both cases of inhomogeneites.
While $W=0$ in the case of inhomogeneous temperature \eqref{inT}, both heat and work are generated in the case of an inhomogeneous Hamiltonian \eqref{inH}. Recalling that both cases of inhomogeneity return us the exact same ensemble-averaged dynamics \eqref{eq:Analytic_Noise}, we note here that their thermodynamical behaviour is in fact significantly different.

\subsection{Work}

\subsubsection{Work generated in a single collision}
\begin{figure*}
\centering
\includegraphics[width=\textwidth]{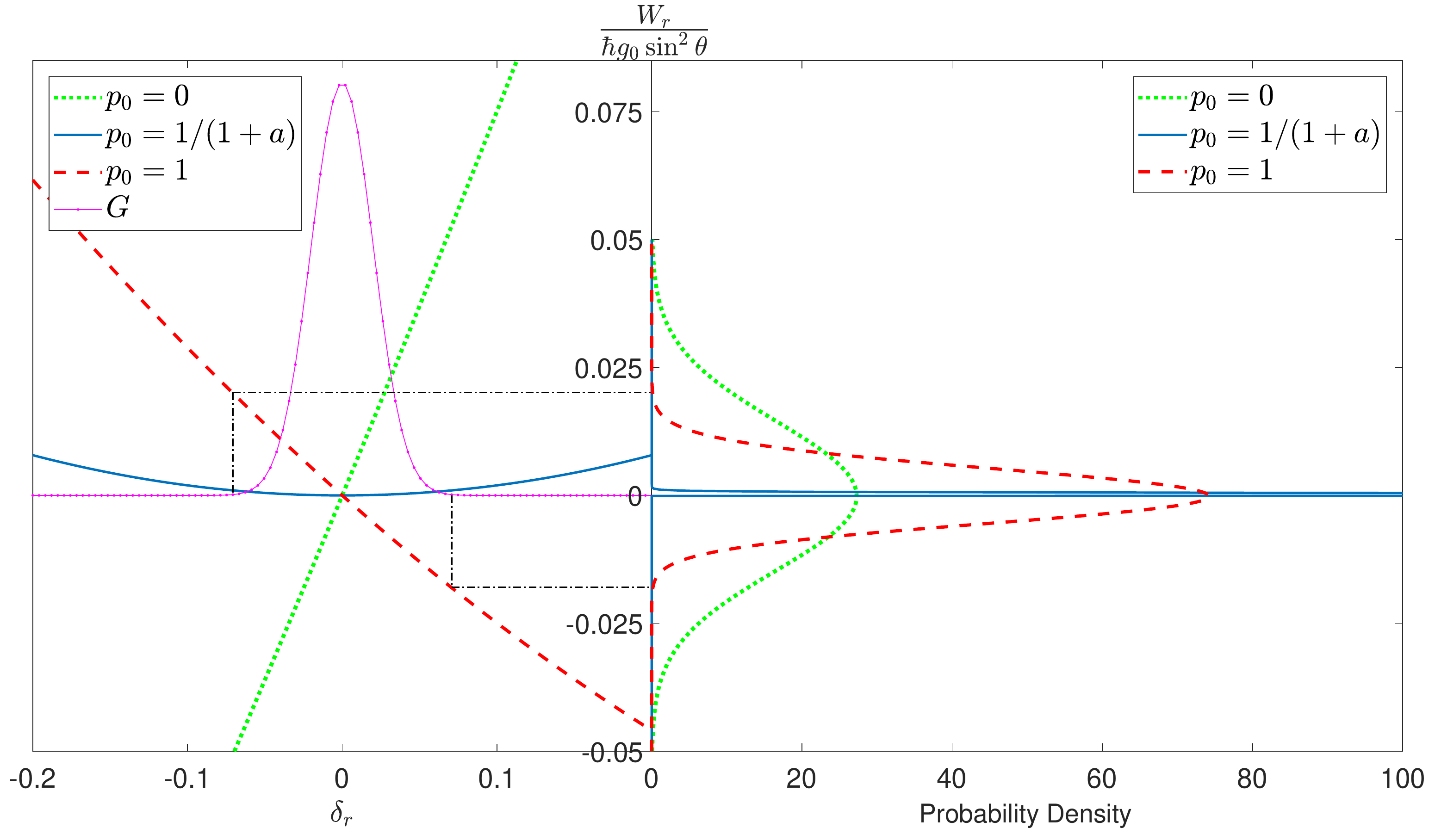}
\caption{(Color online) Single-interaction work and its distribution for qubits, assuming a diagonal input state for the system and an inhomogeneous Hamiltonian. In the left graph, the blue solid, red dashed and green dotted lines are single-interaction work $W_{r}/\hbar g_0\sin^2\theta$ according to Eq.~\eqref{work1} as a function of the inhomogeneity $\delta_{r}$, for various values of $p_0$. The thin magenta line with dot markers shows the distribution Gaussian $G(\delta)$ with $\sqrt{\overline{\delta^2}}=0.02$ that was used for the right plot. The right figure plots the distribution of work $G_W(y)$, with $y=W/\hbar g_0\sin^2\theta$, normalised according to $\int G_W(y)dy=1$, for the same values of $p_0$. Both plots use $\beta\hbar g_0=1$, (whence $a=e^{-1}$ i.e. $\frac{1}{1+a}\approx 0.73$).}\label{figwork}
\end{figure*}

We consider first a single collision between the system and the $r$-th reservoir qudit. The work generated during this collision is \footnote{For simplicity, we have called this work generated in a single collision. Note however, that this is not single-shot work. Instead it would be the average work over an ensemble of frozen inhomogeneities. This term is distinguished from the ensemble average $\overline{W_r}$ in the next subsection which is the actual average over our ensemble whereby the $\delta_{r}$ is drawn independently for each run. Similar distinctions hold for $Q_{r}$. Note that the statistical distribution in Fig.~\ref{figwork} is merely the distribution over the inhomogeneities and not over the fluctuations due to inherent quantum uncertainties.}
\begin{align}
    W_{r}&=\mathrm{Tr}\left[\{U_{S,r}\rho_{r-1}U_{S,r}^{\dagger}-\rho_{r-1}\}H_{\delta_r}\right]
\end{align}
where $\rho_{r-1}=\rho_{S|r-1}\otimes\tau(\delta)$ and $H_{\delta_r}=g_0[s_z^{(S)}+(1+\delta_{r})s_z^{(r)}]$. The calculation eventually yields
\ba\label{workr}
W_{r}&=& {\hbar g_0\, \delta_r \sin^{2}\theta \sum_j j \left[ p_{j}(\underline{\delta}_{r-1}) - q_{j}(\delta_r) \right] } 
\ea 
For qubits, Eq.~\eqref{workr} becomes
\ba\label{work1}
W_{r}&=& \hbar g_0 \, \delta_r \sin^{2}\theta\,[q_{0}(\delta_r)-p_{0}(\underline{\delta}_{r-1})]\;.
\ea Figure \ref{figwork}(left) shows how $W_r$ varies with $\delta_r$ for various values of $p_0$ for qubits. From \eqref{work1} and the knowledge of $G(\delta)$, we can find the \textit{statistical distribution} of single-collision work for qubits. This is easily derived by rewriting \eqref{work1} as
\ba\label{yx}
y&=&\delta\,[q_{0}(\delta)-p_{0}]
\ea with $y\equiv \frac{W}{\hbar g_0\sin^{2}\theta}$. We then invert this function to find the distribution of work $G_W(y)$ induced by the distribution of the inhomogeneity $G(\delta)$. It's clear that the function cannot be inverted analytically. However, we can resort to the Taylor expansion $q_0(\delta)=q_0(0)+{q}'_0(0)\delta +O(\delta^2)$ where $q_0(0)=\frac{1}{1+a}\,,\;\; {q}'_0(0)=\beta\hbar g_{0} \frac{a}{(1+a)^2},$ with $a=e^{-\beta\hbar g_{0}}$. This reduces \eqref{yx} to a quadratic equation
\ba\label{W_dist_quad}
q_{0}'\delta^{2}+(q_{0}-p_{0})\delta-y=0
\ea 
where the notation $q_{0}(0)$ and the like has been shortened to $q_{0}$ for simplicity. From \eqref{W_dist_quad}, the expression of $\delta(y)$ can be easily obtained.

The distribution of $W$ is then given by
\ba\label{gy}
G_W(y)&=&\sum_{s=\pm} G(x_s(y))\,\left|\frac{d\delta_s}{dy'}(y'=y)\right|\nonumber\\
&=&\frac{1}{\sqrt{D}}\sum_{s=\pm} G(\delta_s(y))\,.
\ea where $D=q_{0}'^2+4y(q_0-p_0)$ is the discriminant of equation \eqref{W_dist_quad}. Figure \ref{figwork} (right) illustrates this distribution for a Gaussian distribution of inhomogeneities $G(\delta)$ and a few values of $p_0$. For $p_0=1/(1+a)$, that is for $\rho_{S|r-1}=\tau_S$, the distribution is the narrowest and diverges as $1/\sqrt{W}$ at $W=0$. While the spread of the distribution depends on the input state, the peaks (which coincide with the ensemble averaged work, discussed in the next subsection) do not.

\subsubsection{Ensemble average of single-collision work}\label{EnAvW}

Now we compute the ensemble average $\overline{W_r}$ of \eqref{workr}. One could think that $[H_S+\overline{H}_R,U]=0$ implies $\overline{W_r}= 0$. But this is not the case, because the reservoir states also depend on $\underline{\delta}_N$. The actual expression is 
\begin{align}
\overline{W_r}=-\hbar g_0\, \sin^{2}\theta\, \sum_{j}j\, \overline{\delta_r\,q_{j}(\delta_r)}, 
\end{align} having noticed that $\underline{\delta}_{r-1}$ and $\delta_r$ are not correlated and recalling that our distribution $G(\delta)$ is centered at $\overline{\delta}=0$.

Narrowing our focus to a symmetrical distribution [$G(\delta)=G(-\delta)$], for small $\delta_{r}$, we can make the following Taylor approximation
$q_{j}(\delta_{r})=q_{j}(0)+{q}'_{j}(0)\delta_{r}+{q}''_{j}(0)\delta_{r}^{2}+O(\delta_{r}^{3})$ to find 
\begin{align}
\overline{W_r}&=-\overline{\delta^2} \hbar g_0 \sin^2\theta \sum_{j}jq_{j}'(0)\,\,+\,O(\overline{\delta^4})\,. \label{deltaer}
\end{align} Thus we can conclude that at every step the ensemble average of single-collision work is identical as it is independent of $r$.

\subsubsection{Accumulated work and dynamics}
The work accumulated during the $N$ collisions is \ba\overline{W}_N\,=-\,\sum_{r}\overline{W_r}&\,\approx\,&N \hbar g_0 \overline{\delta^2} \,\sin^2\theta\sum_{j}jq_{j}'(0)\,. \ea 
This may be kept bounded for all $N$ by choosing a suitable scaling of $\theta$ with $N$. However, the value of $\theta$ affects also the dynamics \eqref{eq:Analytic_Noise}: in particular,
\ba
\mathcal{D}(\overline{\rho}_{S|N},\overline{\tau})&=&\cos^{2N}\theta\,\mathcal{D}(\rho_0,\overline{\tau})\,.
\ea where $\mathcal{D}(\rho,\rho')=\frac{1}{2} \textrm{Tr}(|\rho-\rho'|)$ is the trace distance.

Let's then look at the scaling $\sin^2\theta=cN^{-\xi}$. If $\xi>1$, in the limit of large $N$ one has $\overline{W}_N\rightarrow 0$, but also $\mathcal{D}(\overline{\rho}_{N},\overline{\tau})\approx \mathcal{D}(\rho_0,\overline{\tau})$: no work is produced because the dynamics is frozen. If $\xi<1$, then in the limit of large $N$ one has $\mathcal{D}(\overline{\rho}_{N},\overline{\tau})\rightarrow 0$ but $\overline{W}_N\rightarrow \infty$. A good compromise is
\ba
\sin^2\theta=\frac{c}{N}&\Longrightarrow&\left\{
\begin{array}{l}
\overline{W}_N\,\approx\hbar g_0\, \overline{\delta^2}\;c \sum_{j}jq'_{j}(0)\\
\mathcal{D}(\overline{\rho}_{N},\overline{\tau})\approx e^{-c}\,\mathcal{D}(\rho_0,\overline{\tau})
\end{array}
\right.
\ea
The trace distance with the steady state decreases exponentially with $c$, while the total accumulated work increases linearly with $c$ but remains bounded.

\subsection{Heat}
\subsubsection{Heat generation in a single collision}

\begin{figure*}
\centering
\includegraphics[width=\textwidth]{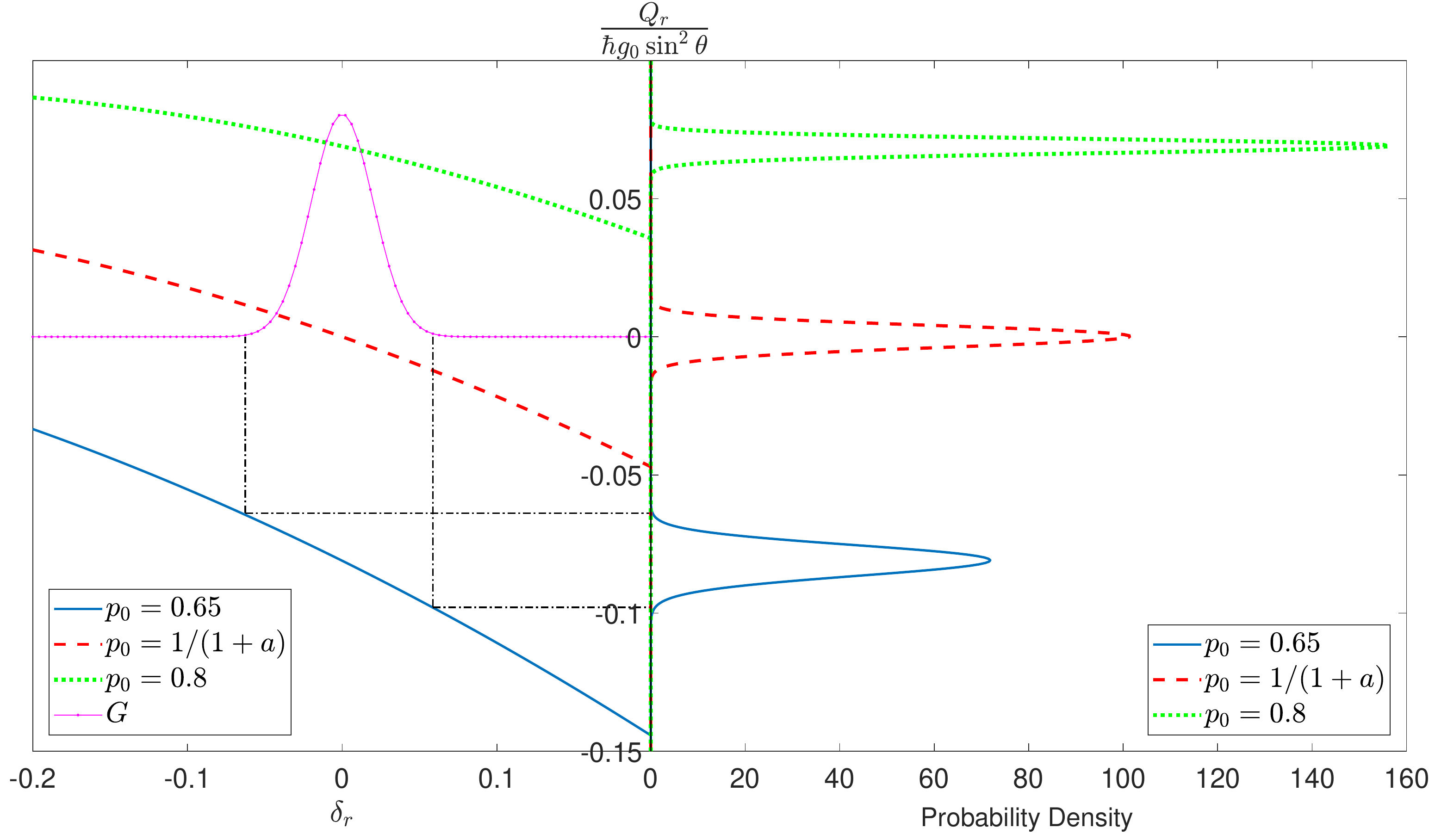}
\caption{(Color online) Single-interaction heat and its distribution for qubits, assuming a diagonal input state for the system and inhomogeneous Hamiltonian. In the left graph, the blue solid, red dashed and green dotted lines are heat generated from a single-interaction $Q/\hbar g_0\sin^2\theta$ according to Eq.~\eqref{heat1} as a function of the inhomogeneity $\delta_{r}$, for various values of $p_0$. The thin magenta line with dot markers shows the Gaussian distribution $G(\delta)$ with  $\sqrt{\overline{\delta^2}}=0.02$ used for the right plot. The right figure plots the distribution of heat $G_Q(y)$, with $y=Q/\hbar g_0\sin^2\theta$, normalised according to $\int G_Q(y)dy=1$, for the same values of $p_0$. Both plots use $\beta\hbar g_0=1$, (whence $a=e^{-1}$ i.e. $\frac{1}{1+a}\approx 0.73$). }\label{figheatstats_inH}
\end{figure*}

We first limit our attention to one single collision between the system and the $r$-th reservoir qudit. The heat generated in this collision is
\begin{align}
    Q_{r}&=\mathrm{Tr}\left[\{\tau(\delta_r)-\rho_{R|r}\}\,g_0(1+\delta_r) s_z^{(r)}\right]
\end{align}
where $\rho_{R|r}=\mathrm{Tr}_{S}[U_{S,r}\{\rho_{S|r-1}\otimes\tau(\delta_r)\}U_{S,r}^{\dagger}]$ is the partial state of the $r$-th reservoir qudit after one application of the unitary interaction $U_{S,r}$ on the system and the reservoir. This calculation returns

\begin{align}\label{heatr}
Q_{r}=&\hbar g_{0}(1+\delta_r)\sin^2\theta\times\nonumber\\
&\sum_{j}\left(j-\frac{d-1}{2}\right)\left[q_{j}(\delta_r)-p_{j}(\underline{\delta}_{r-1})\right].
\end{align} For qubits, Eq.~\eqref{heatr} reduces to
\begin{align}\label{heat1}
Q_{r}&=&\hbar g_0 (1+\delta_r)\sin^{2}\theta [p_0(\underline {\delta}_{r-1})-q_0(\delta_r)]\,
\end{align}
for which, like the case for work, we can calculate the statistics of single-collision heat (see Fig.~\ref{figheatstats_inH}).

With the Taylor expansion, we will obtain the following quadratic equation for \eqref{heat1}
\ba
q_{0}'\delta^{2}+(q'_{0}+q_{0}-p_0)\delta+(y-p_0+q_0)=0\,,
\ea
with $y=\frac{Q_{r}}{\hbar g_{0}\sin^{2}\theta}$ and as previously, $q_{0}$ is shorthand for $q_{0}(0)$. From this, we can again obtain $\delta$ in terms of $y$, and the statistics of heat is then given by \eqref{gy}.

Unlike the case of work however, the distribution (Fig.~\ref{figheatstats_inH} [right]) is \textit{not} the narrowest for a state that is close to the thermal state $p_0=1/(1+a)$, but is in general narrower for input states that have lower energy. We find also that the peak of these distributions depend on the state of the system which is sensible as heat is the amount of energy the reservoir dumps into the system. Therefore the average $\overline{Q_{r}}$ (peak of the distribution) in this partial swap model necessarily depends on the energy of the system interacting with the reservoir.

\subsubsection{Ensemble average of single-collision heat}
Turning now to the ensemble average $\overline{Q_r}$ of Eq.~\eqref{heatr}, we obtain the actual expression for $\overline{Q_r}=\hbar g_0\sin^2\theta\sum_{j} \left(j-\frac{d-1}{2}\right)\left[\overline{q_j(\delta_r)}+\overline{\delta_r q_j(\delta_r)}-\overline{p_j(\underline{\delta}_{r-1}})\right]$. Note again the independence of $\delta_r$ and $\underline{\delta}_{r-1}$ that allows us to split the averages, and the chosen distribution allows us to drop terms proportional to $\overline{\delta}$.

In a similar fashion to the case of work, we consider the Taylor expansion of $q_{j}(\delta)$ to find
\begin{align}
\overline{Q_{r}}=\hbar g_0\sin^2\theta \Sigma_{Q_r}\bigl(\overline{\delta^2}\bigr)+O(\overline{\delta^{4}})
\end{align} 
where 
\begin{align*}
\Sigma_{Q_r}\bigl(\overline{\delta^{2}}\bigr)=\sum_{j}&\left(j-\frac{d-1}{2}\right)\times\\
&\left[q_j(0)-\overline{p_{j}(\underline{\delta}_{r-1}})+\left\{q'_{j}(0)+q''_{j}(0)\right\}\overline{\delta^{2}}\right]
\end{align*}
%For qubits, we can simplify it to
%\begin{align}
%\overline{Q_r}=&\hbar g_0\sin^2\theta\times\nonumber\\
%&\left[\overline{p_0(\underline{\delta}_{r-1})}-q_{0}(0)-\left\{q_{0}'(0)+q_{0}''(0)\right\}\overline{\delta^{2}}\right]
%\end{align}
Unlike the expression of work however, we notice that $\overline{Q_r}$ depends on $r$. 

A quick calculation for the energy of the system returns us
\begin{align}\label{intE}
\Delta U_{r}=\hbar g_{0}\sin^{2}\theta\sum_{j}&\left(j-\frac{d-1}{2}\right)\times\nonumber\\
&\left[q_{j}(\delta_r)-p_{j}(\underline{\delta}_{r-1})\right]
\end{align}
for the single collision, and for the ensemble average utilizing Taylor expansion we have,
\begin{align}
\overline{\Delta U_{r}}=&\hbar g_{0}
\sin^{2}\theta\sum_{j}\left(j-\frac{d-1}{2}\right)\times\nonumber\\
&\left[q_{j}(0)-\overline{p_{j}(\underline{\delta}_{r-1})}+q''_{j}(0)\overline{\delta^{2}}\right]+O(\overline{\delta^{4}})
\end{align}
Recalling that the probability is normalized, and $\sum_{i}q_{i}'(0)=0$ for  the first order as well as all higher order derivatives, one obtains the first law, as expected.

%\footnote{When the Hamiltonian is time-dependent, work is usually defined as $W(t_2,t_1)=\int_{t_1}^{t_2} \textrm{Tr}(\rho(t)\dot{H}(t))dt$. In our model, a collision can be seen in this framework: the interaction Hamiltonian $H_{int}$ that generates $U$ is switched on abruptly at $t_1$, then switched off abruptly at $t_2=t_1+t_{int}$. Thus $\dot{H}_{int}(t)=H_{int}[\delta(t-t_1)-\delta(t-t_2)]$. Noticing that during the interaction $H_S+H_R+H_{int}$ is obviously conserved, we find indeed $W(t_2,t_1)=\textrm{Tr}[(\rho(t_2)-\rho(t_1))(H_S+H_R)]$.}. We focus on the case \eqref{inH} of inhomogeneous Hamiltonian, because in the case \eqref{inT} it holds $[H_S+H_R,U]=0$ and no work is generated during any collision.

%***************************************************************************************************************************************
%*************************** Section III ****************************
%***************************************************************************************************************************************

\begin{figure*}
\centering
\includegraphics[width=\textwidth]{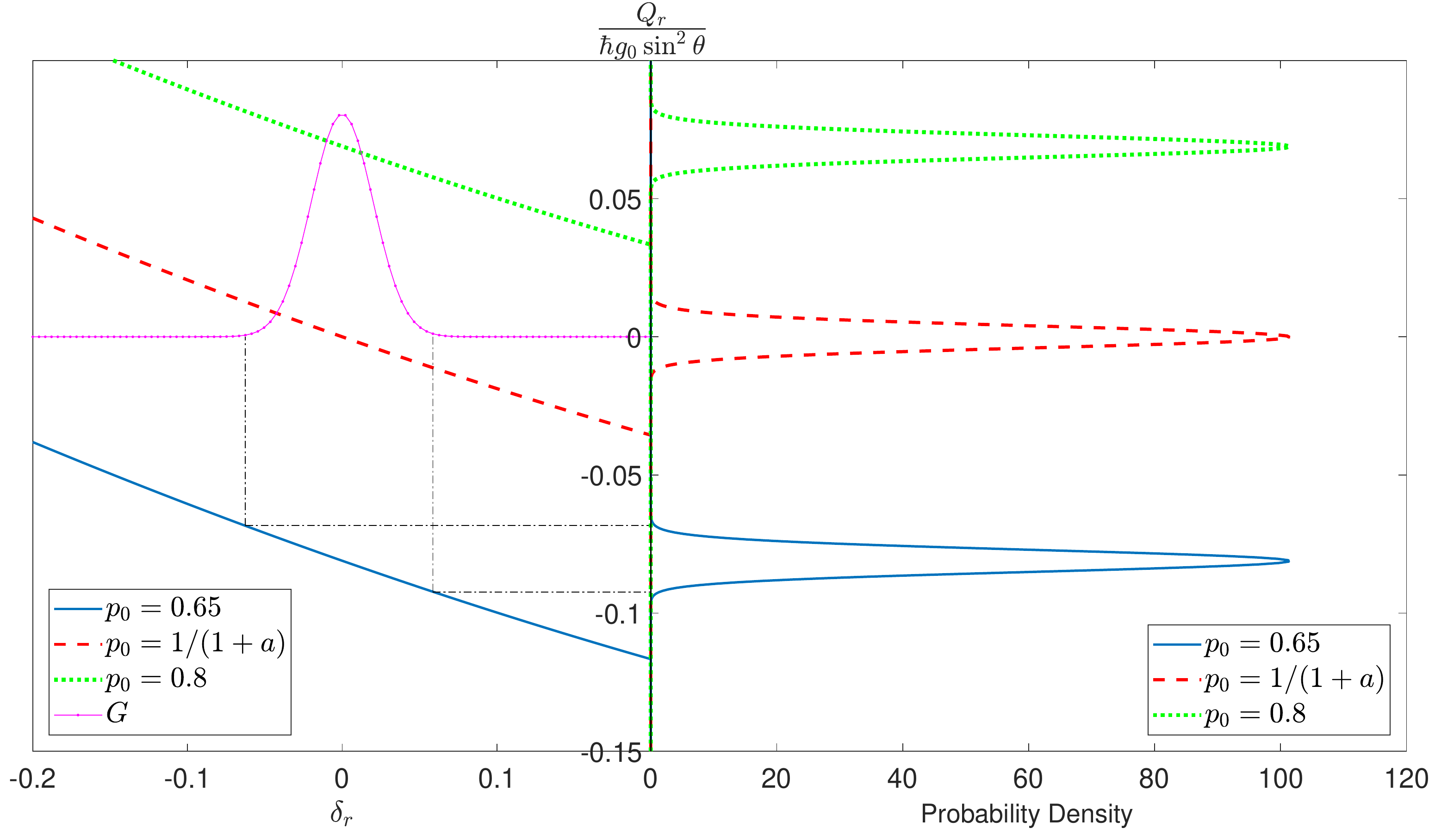}
\caption{(Color online) Single-interaction heat and its distribution for qubits, assuming a diagonal input state for the system and inhomogeneous temperature. In the left graph, the blue solid, red dashed and green dotted lines are heat generated from a single-interaction $Q/\hbar g_0\sin^2\theta$ according to Eq.~\eqref{heat2} as a function of the inhomogeneity $\delta_{r}$, for various values of $p_0$. The thin magenta line with dot markers shows the Gaussian distribution $G(\delta)$ with  $\sqrt{\overline{\delta^2}}=0.02$ used for the right plot. The right figure plots the distribution of heat $G_Q(y)$, with $y=Q/\hbar g_0\sin^2\theta$, normalised according to $\int G_Q(y)dy=1$, for the same values of $p_0$. Both plots use $\beta\hbar g_0=1$, (whence $a=e^{-1}$ i.e. $\frac{1}{1+a}\approx 0.73$).}\label{figheatstats_inT}
\end{figure*}

\section{Work, heat and first law -- Inhomogeneous temperature}
\label{ssenergy}
For the inhomogeneous temperature \eqref{inT}, it holds that $[H_S+H_R,U]=0$ and no work is generated during any collision, therefore we only have heat $Q$. Since $W=0$, the first law in the present case of \eqref{inT} is merely $\Delta U=Q$. 

Note further that $\Delta U$ is the same regardless of whether the inhomogeneity is due to fluctuations in the Hamiltonian or the temperature, as $\Delta U$ depends only on the Hamiltonian of the system and the dynamics of the reduced system, which are identical in both cases. Therefore, we already know that $\Delta U$ in this scenario is exactly Eq.~\eqref{intE}. A quick calculation of $Q_{r}$ for returns us Eq.~\eqref{intE} too, as expected. For qubits, 
\begin{align}\label{heat2}
Q_{r}=\hbar g_{0}\sin^{2}\theta \left[p_0\left(\underline{\delta}_{r-1})-q_{0}(\delta_{r}\right)\right]
\end{align}

As in the case of of inhomogeneous Hamiltonian, we can determine the statistics of heat as well (Fig.~\ref{figheatstats_inT}). By the Taylor expansion, we will obtain 
\ba
q_{0}''\delta^{2}+q_{0}'\delta+(q_{0}-p_{0}+y)=0\,,
\ea for \eqref{heat2}, where ${q}''_{0}\equiv{q}''_0(0)=(\beta\hbar g_{0})^{2} \frac{a(a-1)}{(1+a)^3}$ and $y=\frac{Q_{r}}{\hbar g_{0}
\sin^{2}\theta}$. As usual, the distribution of $G_{Q_{r}}$ will be given by the equation \eqref{gy}

We note that unlike the statistics of heat for an inhomogeneous Hamiltonian, the spreads of the distribution here does not depend on the input state.

%***************************************************************************************************************************************
%*************************** Section IV ****************************
%***************************************************************************************************************************************

\section{The ``second laws of thermal operations'' and inhomogeneous reservoirs}
\label{seclaws}

The set of criteria under which a target state $\rho'$ can be obtained from $\rho$ by free evolution can be seen as the analog of the second law of thermodynamics. The transformation $\rho\longrightarrow\rho'$ under free operation does not define a total order: as a result, it cannot be characterised by a single criterion \cite{Gour_2015}. Brand\~{a}o and coworkers \cite{Brandao_2015} wrote the \textit{second laws of thermal operations} as the monotonical decrease \ba\label{laws}
\Delta F_\alpha\,=\,F_{\alpha}(\mathcal{E}[\rho]||\tau_S)-F_{\alpha}(\rho||\tau_{S})\leq 0&,&\alpha\in\mathbb{R}
\ea
of a continuous family of \textit{generalised free energies} \ba F_{\alpha}(\rho||\tau_{S})&=&k_{\rm B}T\, \left[D_{\alpha}(\rho||\tau_{S})- \log Z_S\right]\ea defined from the $\alpha$-R\'enyi divergence $D_{\alpha}(\rho||\tau_{S})$. If $\rho$ and $\tau_{S}$ are diagonal in the same basis, as we are assuming since the beginning, it holds
\begin{align} 
D_{\alpha}(\rho||\tau_{S})=\frac{\text{sgn}(\alpha)}{\alpha-1}\log\sum_{j}p_{j}^{\alpha}q_{j}^{1-\alpha}\label{eq:RenyiDiv}
\end{align} with $q_j=e^{-\beta E_j}/Z_S$ the eigenvalues of $\tau_S$.

The conditions \eqref{laws} are necessary and sufficient for free evolution. Since inhomogeneous reservoirs deviate from free dynamics, they should violate these conditions in some cases. The following protocol leads to a violation for \textit{all} $\alpha$: prepare the system in the state $\tau_S$ and let it evolve to $\bar{\tau}$ according to \eqref{eq:Analytic_Noise}. In this case, $\beta\Delta F_\alpha= D_{\alpha}(\bar{\rho}_{S|N}||\tau_{S})-D_{\alpha}(\tau_S||\tau_{S})$ is strictly positive, since $D_{\alpha}(\rho||\tau_{S})\geq 0$ with equality if and only if $\rho=\tau_S$. 

Updating the laws \eqref{laws} to take into account any deviation from free evolution is an open challenge. Our study of inhomogeneous reservoirs may serve as a starting point for this task. We first stress that, in our model, the possible state transformations are given by \eqref{eq:Analytic_Noise} for both inhomogeneous temperature and Hamiltonian. The generalised laws that single out these transformations must therefore be independent of the type of inhomogeneity \footnote{The replacement of $\tau_S$ with $\overline{\tau}$ in \eqref{laws} is a formal fix that ignores the physics of the problem. With such a fix, every contractive map would define a ``second law'', without any reference to thermodynamics.}. 

\begin{figure}[h]
\centering
\includegraphics[width=0.45\textwidth]{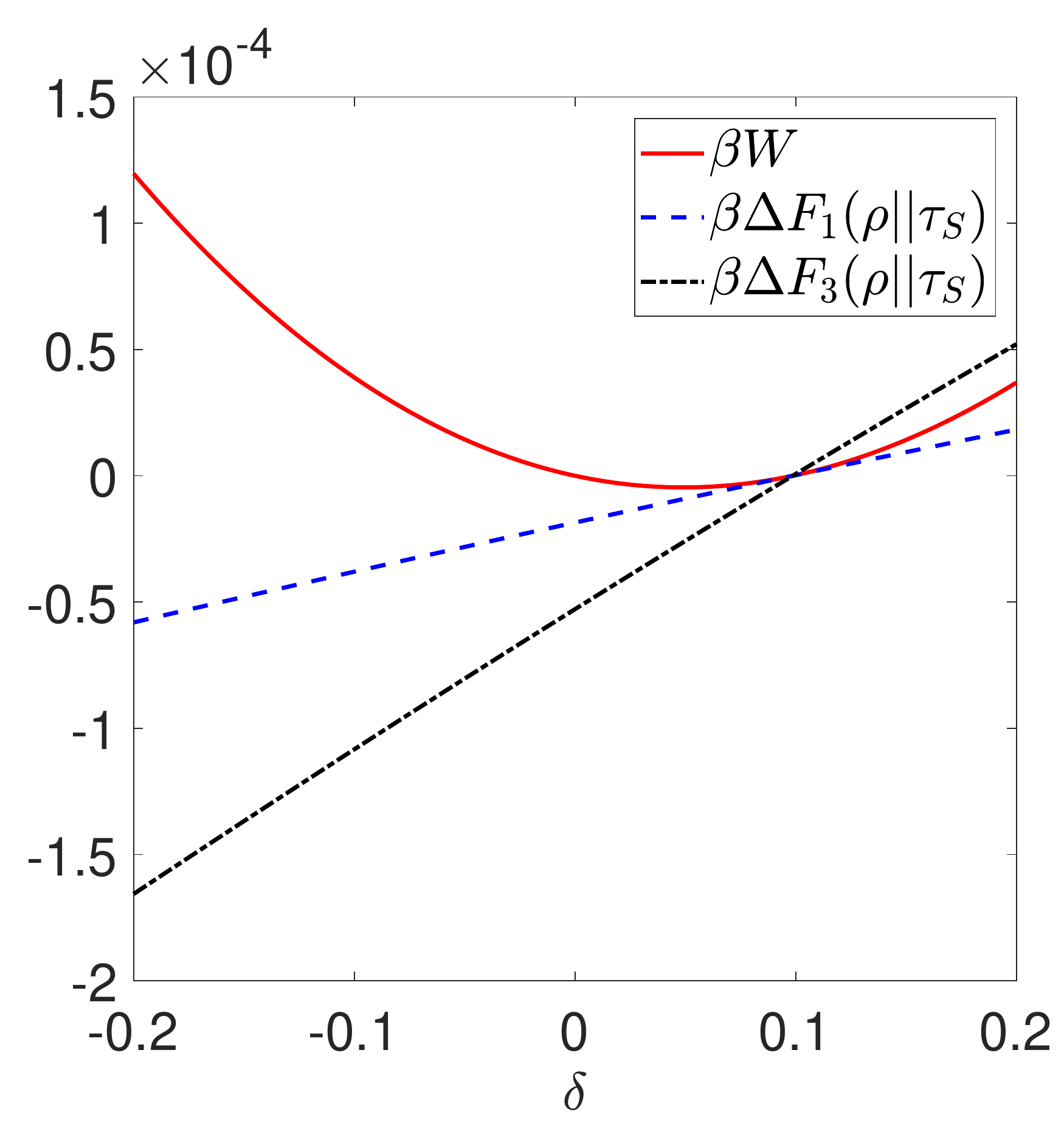}
\caption{(Color online) Comparison of $\beta W$ and $\beta \Delta F_\alpha$ after a single interaction (mixing angle $\theta=0.1$ rads) as a function of $\delta$, for $\hbar g_0=1$  and input state characterised by $p_0=0.75$. The expected violation of conditions \eqref{laws} happens for $\delta>0$. $\beta \Delta F_1$ (blue dashed) is upper bounded by $\beta W$ (red solid), as it should; but for larger values of $\alpha$, this upper bound is also violated (plotted for $\alpha=3$ [black dash-dotted]).}\label{figalpha}
\end{figure}

However, their thermodynamical meaning will have to be different. When work is generated and $\beta$ is unique, thermodynamics requires $\Delta F_1\leq W$, which was indeed proved for collisional models \cite{Strasberg_2017}, and holds true for our model as well. Our model of inhomogeneous Hamiltonian \eqref{inH} shows that a generalisation $\Delta F_\alpha\leq W$ won't hold for $\alpha>1$ \footnote{The bound $\Delta F_\alpha\leq W$ was obtained for all $\alpha$ in Section G.3 of the Supplementary Information of \cite{Brandao_2015}. But their definition of work is different: they are looking at state transformations catalysed by a two-level battery (a ``work bit'') prepared in the thermal state, and $W$ is the value of the gap. In other words, $W$ is a parameter of the state, chosen so that the state transformation becomes possible, and is not related to the time-dependent dynamics (also, its value does not match the ``change in energy'' of the joint system).}, see Figure \ref{figalpha}. In the case of inhomogeneous temperature \eqref{inT}, work is not generated; and in fact, in this narrative, the laws should not even involve free energies, since the second law of thermodynamics can be cast in terms of free energy only if the system is in contact with a bath at a single temperature. One could opt for reading \eqref{inT} in the narrative of resource theories, where there is still a single reference temperature $\beta$, the $\tau(\delta_r)$ playing the role of non-thermal (i.e.~resource) states. In this context, Ref.~\cite{Meer_2017} defined approximate second laws with free energies $F_\alpha^\varepsilon$ where $\varepsilon$ is the maximal distance between a target state reachable with free operation and one reachable with the resource operation. In our case $\varepsilon=\mathcal{D}(\tau_S,\overline{\tau})$. For an analytical estimate for qubits, we compute the upper bound $\varepsilon\lesssim \overline{\mathcal{D}(\tau_S,\tau(\delta))}=\sqrt{2/\pi}\beta g_{0}\hbar\frac{a}{(1+a)^2}\,\sqrt{\overline{\delta^{2}}}+O(\delta^{2})$.

There are indeed other alternatives to the second law \cite{Esposito_2010, Reeb_2014}, and it can be easily verified that the entropy production $\Delta S_{r}=D[\rho_r||\rho_{S|r}\otimes\tau(\delta_{r})]$ in \cite{Esposito_2010} is always positive, and hence, this alternative second law is always obeyed. $\rho_{r}$ here is the combined state of the system and one reservoir qudit after one interaction. Here however, one can no longer speak of a family of necessary and sufficient conditions for a particular evolution. In the interest of understanding how these family of necessary and sufficient conditions relax in the presence of small inhomogeneities, updating the laws \eqref{laws} remains an open challenge.

\section{Long-term Behaviour}
In the preceding sections, we have only discussed the thermodynamic behaviour for a single step. One could also be interested in the thermodynamic behaviour of the system over many steps as the system thermalizes. The pertinent point of query here is as follows: in the long run, how robust are these second laws with respect to small inhomogeneities in the reservoir? Do they deviate significantly in the presence of small inhomogeneities? 

\begin{figure}[h]
\centering
\includegraphics[width=0.45\textwidth]{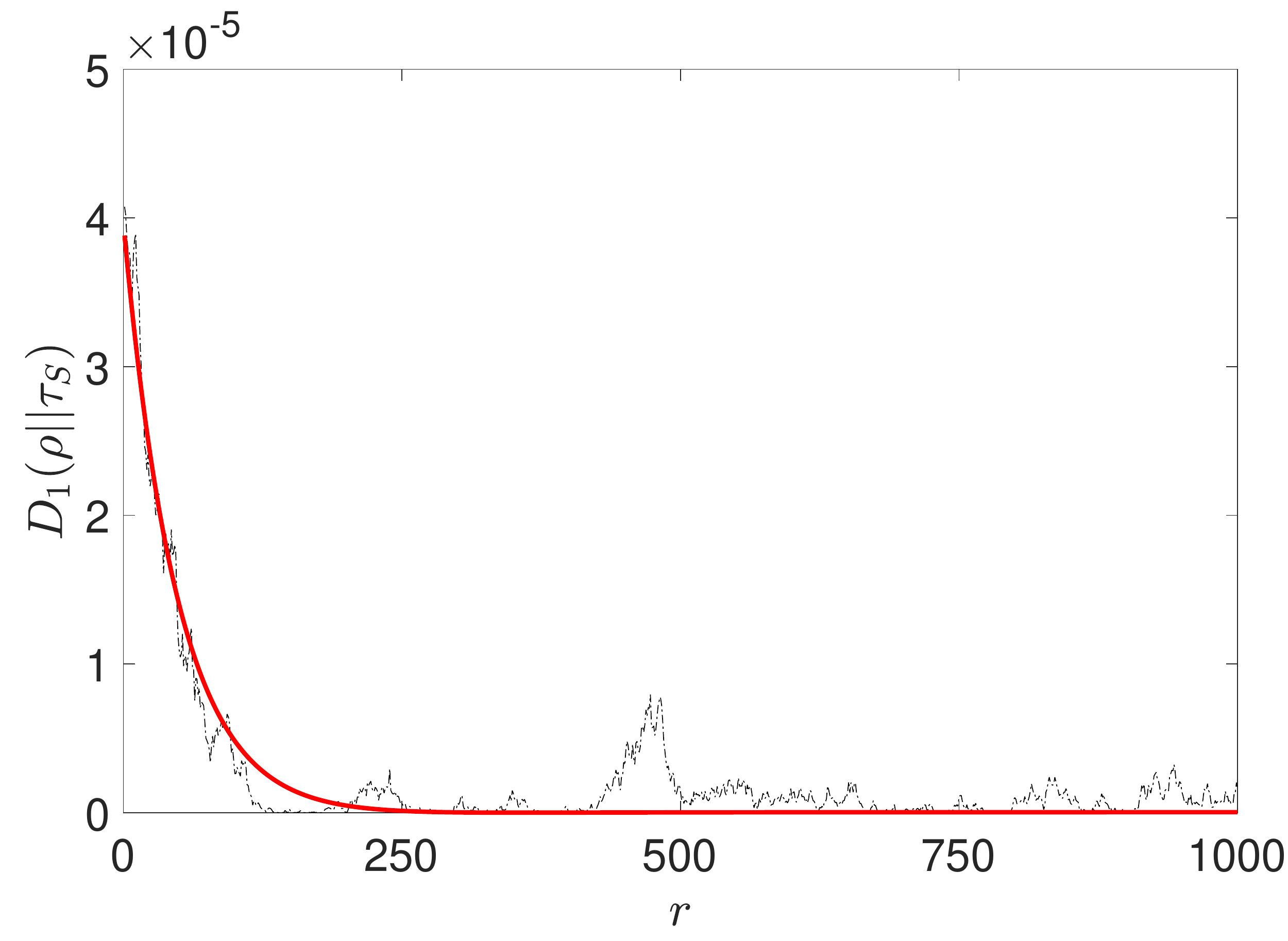}
\includegraphics[width=0.45\textwidth]{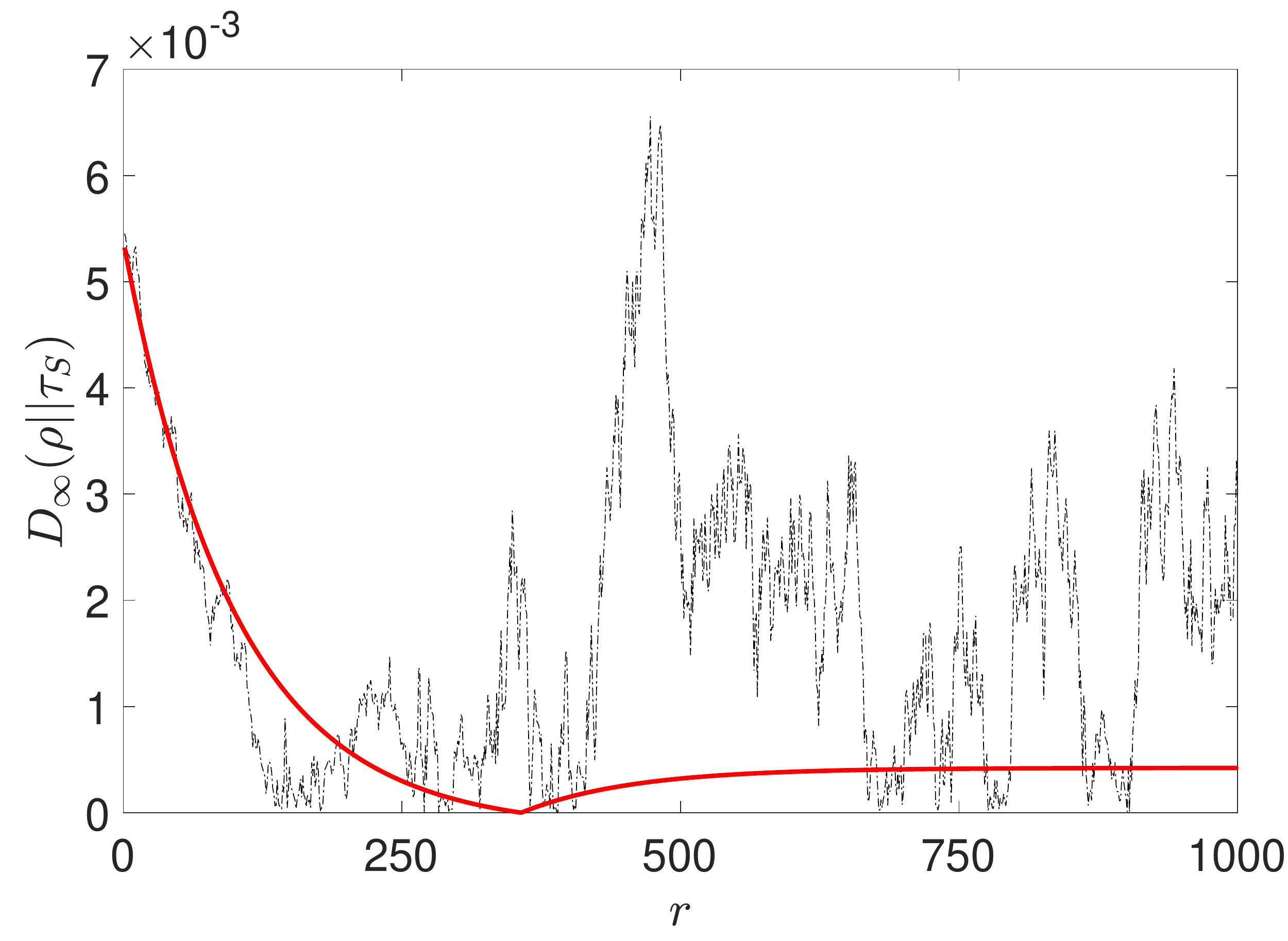}
\caption{(Color online) Comparison of the R\'{e}nyi divergences for $\alpha=1$ and $\alpha=\infty$. The black dash-dotted plots are values of $D_{\alpha}(\rho_{S|r}||\tau_{S})$ over a frozen ensemble and the red solid plots are the free energy calculated for the ensemble averaged state $D_{\alpha}(\overline{\rho}_{S|r}||\tau_S)$ for each step. For both plots, $\beta \hbar g_0=1$, $\sqrt{\overline{\delta^2}}=0.05$ and input state characterised by $p_0=0.735$.}\label{figLongTerm}
\end{figure}

We note that whilst the free energies for lower $\alpha$ values are relatively robust, the free energies for high $\alpha$ are indeed very unstable in the presence of small inhomogeneities. In Fig.~\ref{figLongTerm}, we plot our numerical simulations of both the R\'{e}nyi divergence over a frozen ensemble $D_{\alpha}(\rho_{S|r}||\tau_S)$ as well as that of the ensemble average state $D_{\alpha}(\overline{\rho}_{S|r}||\tau_{S})$. The plots are of the R\'{e}nyi divergence instead of the free energies as the shape of the graph is unaffected by this choice.

\section{Conclusion}
\label{sec:conclusion}

Extending the resource theory of thermal operations to non-ideal reservoirs is not trivial \cite{Sparaciari_2017,Scharlau2018,Richens2018}. In this paper, we have introduced the notion of inhomogeneous reservoirs. Using the most standard collisional model, which fits well the definition of free dynamics in the absence of inhomogeneity, we have studied the two simplest cases of i.i.d.~inhomogeneities: either in local temperature (which can be interpreted as having ``resource states'') or in the local Hamiltonian (which is an instance of ``resource operations''). These two notions of inhomogeneity have a clear physical flavour and both predict the exact same dynamics. However, we note that the thermodynamic behaviour of the system differs significantly. Furthermore, we note that while the lower $\alpha$ free energies are somewhat more robust, the higher $\alpha$ free energies are very sensitive to these inhomogeneities.

There are clearly many ways in which this study can be extended. Here we have restricted our attention to states of the system that are diagonal in the energy eigenbasis, and it would be worth considering general states of the systems and the role of coherence. Also, even staying within the family of collisional models, one can study different parameters. A standing open problem is the formulation of the rules for state transformation (``second laws'') for inhomogeneous reservoirs: this paper has provided only an initial insight on this question. 

%***************************************************************

\section*{acknowledgments}

We acknowledge illuminating discussions with and useful feedback from Alvaro Alhambra, Philippe Faist, Rodrigo Gallego, Gabriel Landi, Matteo Lostaglio, Kavan Modi, Markus M\"uller, Nelly Ng and Henrik Wilming.

This research is supported by the National Research Fund and the Ministry of Education, Singapore, under the Research Centres of Excellence programme.

\bibliography{refs}

\end{document}